\documentclass[12pt,a4paper]{article}
\usepackage{graphicx} 
\begin{document}
\title{Microscopic calculation of proton capture reactions in mass 60-80 region and its astrophysical 
implications}
\author{Chirashree Lahiri and G. Gangopadhyay$^1$\\
Department of Physics, University of Calcutta\\ 92 A.P.C. 
        Road, Kolkata 700009, India\\
$^1$e-mail : ggphy@caluniv.ac.in}  

\maketitle
\abstract{
Microscopic optical potentials obtained by folding the DDM3Y 
interaction with the densities from Relativistic Mean Field approach have 
been utilized to evaluate S-factors of low-energy $(p,\gamma)$ reactions 
in mass 60-80 region and to compare with experiments. The Lagrangian density 
FSU Gold has been employed. Astrophysical rates for important  proton capture 
reactions have been calculated to study the behaviour of rapid proton 
nucleosynthesis for waiting point nuclei
with mass less than $A=80$.}
\date{}

\section{Introduction}
Relativistic Mean Field (RMF) calculations have proved to be very successful in 
describing different features of nuclei. This  
method has been used to study binding energy of the ground state and various 
excited states, deformation, charge radii, density profile and nuclear halo, 
etc\cite{RMF1}. Particularly, the success in reproducing the density has 
motivated various microscopic calculations that generate nucleon-nucleus and
nucleus-nucleus potentials for study of proton and alpha radioactivity and 
scattering (See \cite{decay,NiCu,CBe} and Refs. therein). In the present work, 
we use RMF densities to produce microscopic optical potentials\cite{mom} to 
study proton capture reactions in mass 60-80 region. 

Proton capture reactions at very low energy play a very important role in
nucleosynthesis. Particularly rapid proton capture ($rp$) process in explosive 
nucleosynthesis is a basic ingredient in driving the abundance along the 
$N=Z$ line\cite{book,rep}. As this process has to overcome a large Coulomb barrier 
it can occur only at a higher temperature range. For example, X-ray bursts 
provide a large flux of protons at peak temperatures around 1-2 GK and are 
expected to play a significant role in the creation of nuclei up to mass 110.

The $rp$-process proceeds along the $N=Z$ line in mass 60-80 region. In nature, 
the important proton capture reactions usually involve certain nuclei as 
targets which are not available to us. Hence, experimental information is 
difficult, if not impossible, to obtain, at least in near future. In such a 
situation, one has to rely on theory for the reaction rates. 
Rauscher {\em et al.} have extensively calculated reaction rates and cross 
sections in a 
global approach\cite{astro,rate1}. They have commented that statistical model 
calculations may be improved by using locally tuned parametrization of nuclear 
properties such as optical potential. However, they prefer a global approach 
to predict astrophysical rates for experimentally inaccessible nuclei.
Cross sections have been calculated for proton capture reactions in mass 60-80 
regions using a semimicroscopic optical potential in the local density
approximation using phenomenological density prescriptions. However, far from 
the stability valley, these prescriptions may not represent the actual 
densities very well leading to considerable uncertainty in the reaction 
rates. Very often, the reactions rates are varied by a large factor to 
study their effect. For example, Schatz\cite{old} varied the rates of certain 
reactions by a factor of hundred. Obviously, this makes the results
uncertain to some extent.

A fully microscopic calculation may be used to estimate the rates to reduce 
the above uncertainty. A consistent framework for calculation may be 
constructed based on microscopic densities. This has the advantage of extending 
it to  unknown mass regions. In the present work, we have tried to calculate 
the reaction rates from a purely microscopic model, {\em i.e.} RMF. 
We have already mentioned that RMF is particularly suitable to describe nuclei 
far away from the stability valley where experimental information is scarce.
A microscopic optical potential obtained by folding an appropriate 
microscopic NN interaction is expected to be more accurate and may do away
with the necessity of any arbitrary variation in the reaction rates.

However, even in microscopic optical model, 
there often remain certain parameters which can be fixed only after comparison 
with experiment. We have compared the results for a number of reactions in the 
mass region $A=60-80$ for which experimental informations are available. This has
helped us in determining a set of parameters for this mass region.

Once the parameters have been fixed, we employ them to calculate the rates of 
a number of astrophysically important proton capture reactions.
Certain $N=Z$ nuclei having the highest abundance in an equilibrium in a chain 
are called the waiting points\cite{book} for the chain. These nuclei have 
negative or small positive Q-values for proton capture. An equilibrium between 
the $(p,\gamma)/(\gamma,p)$ processes is established and the $rp$ process may 
have to wait for beta-decay or $\alpha$-capture to proceed to heavier nuclei. 
Certain $N=Z$ waiting point nuclei with $A<80$, {\em viz.} $^{64}$Zn,
$^{68}$Se, $^{72}$Kr, and $^{76}$Sr have long half lives, their total lifetime 
being large compared to the time scale of typical X-ray bursts (10-100 sec). 
Thus, they may produce a bottleneck in the 
$rp$-process that would slow down the rate of hydrogen burning and necessitate 
extended burst tails unless two proton capture can reduce these half lives
and bridge the waiting points. 
X-ray burst model calculations are therefore particularly sensitive to the 
rates of proton capture for these nuclei. We have used the microscopic approach,
used in the present work, to calculate the rates with an aim to study the 
bridging of the waiting point nuclei.

\section{Method}

As already mentioned, experimental $(p,\gamma)$ rates in many nuclei involved 
in the $rp$-process are not available as they are unstable. Hence, theory 
remains our sole guide. A microscopic optical model calculation based on 
theoretical mean field densities are expected to provide us with reliable rates. With this in mind, 
we have calculated the nuclear density profiles in a 
RMF approach using the Lagrangian density FSU 
Gold\cite{fsu}. This density contains two additional non-linear meson-meson interaction 
terms, whose main virtue is a softening of both the 
Equation of State of symmetric matter and symmetry energy. As a result, the new 
parametrization becomes more effective in reproducing quite a few nuclear 
collective modes, namely the breathing modes in $^{99}$Zr and $^{208}$Pb, and 
the isovector giant dipole resonance in $^{208}$Pb\cite{fsu}. We have found 
this Lagrangian density to be very useful in explaining various other properties that depend 
on the nuclear density, such as nuclear decay and reaction\cite{decay,NiCu,CBe}.

As the important quantity in calculating proton capture cross-section is the 
nuclear density as a function of radius, the calculations 
have been carried out in co-ordinate space assuming spherical symmetry. 
Pairing has been introduced under the BCS approximation using a zero range 
pairing force of strength 300 MeV-$fm$ for both protons and neutrons.
The RMF+BCS equations are solved under the usual assumptions of classical 
meson fields, time reversal symmetry, no-sea contribution, etc. 
The details of the calculation may be obtained from Bhattacharya {\em
et al.}\cite{MB,MB1}.

Microscopic optical model potential is usually obtained by folding an  
effective interaction, derived from the nuclear matter calculation, in the 
local density approximation, {\em i.e.} by substituting the nuclear matter 
density with the density distribution of the finite nucleus.  
In the present work, the  density dependent M3Y (DDM3Y) effective 
interaction \cite{ddm3y1,ddm3y2,ddm3y3,ddm3y} has been utilized for this 
purpose. This was obtained from a finite range energy independent G-matrix 
elements
of the Reid potential by adding a zero range energy dependent pseudopotential 
and introducing a density dependent factor.
The interaction is given by
\begin{equation}
 v(r,\rho,E)=t^{M3Y}(r,E)g(\rho)                 
\end{equation} 
where $E$ is incident energy and $\rho$, the nuclear density. 
The $t^{M3Y}$ interaction is given by
\begin{equation}
t^{M3Y}=7999\frac{e^{-4r}}{4r}-2134\frac{e^{-2.5r}}{2.5r}+J_{00}(E)\delta(r)
\end{equation} 
where $J_{00}(E)$ is the zero range pseudo potential, 
\begin{equation}
J_{00}(E)=-276\left( 1-0.005\frac{E}{A}\right) {\rm MeV} fm^{3}\end{equation} 
and $g(\rho)$ the density dependent  factor,
\begin{equation}
g(\rho)=C(1-b\rho^{2/3})\end{equation} 
The constants in the last equation have been obtained from nuclear
matter calculation\cite{ddm3y} as $C=2.07$ and $b=1.624$ $fm^2$.
We have used this form in our calculation keeping the above parameters 
unchanged.

Since nuclear matter-nucleon potential does not include a spin-orbit term, the
spin-orbit potential from the Scheerbaum prescription\cite{SO} coupled with the
phenomenological complex potential depths $\lambda_{vso}$ and $\lambda_{wso}$ 
has been used.
\begin{equation} U^{so}_{n(p)}(r)=(\lambda_{vso} +i\lambda_{wso})
\frac{1}{r}\frac{d}{dr}(\frac{2}{3}\rho_{p(n)}+\frac{1}{3}\rho_{n(p)})
\end{equation}
The depths are functions of energy, given by 
\begin{displaymath}\lambda_{vso}=130\exp(-0.013E)+40\end{displaymath}
\begin{displaymath}\lambda_{wso}=-0.2(E-20)\end{displaymath}
where $E$ is in MeV. These standard values have been used unaltered in the 
present work.

Cross-section and astrophysical rates are  calculated in the Hauser-Feshbach 
formalism using the computer package TALYS1.2\cite{talys}. The present method 
has already been applied in mass 60 region in Ref.\cite{NiCu}. In the present 
approach, though, a slightly different normalization has been applied as 
described later. 

For the calculation of proton capture at waiting points,  a small network has 
been designed which includes the following processes. The waiting point nucleus with 
$Z=N$, which acts as a seed, may capture a proton. The resulting nucleus, with 
$Z=N+1$, may either capture another proton or undergo photodisintegration 
emitting a proton to go back to the seed nucleus. The nucleus with $Z=N+2$ may 
also undergo photodisintegration. In addition, all the three nuclei mentioned 
above may undergo $\beta$-decay. The photodisintegration rates at different 
temperatures have been calculated from the proton capture rates using the 
principle of detailed balance. The density has been taken as $10^6$ gm/cm$^3$
unless otherwise mentioned. The proton fraction has been assumed to be 0.7. 

One of the difficulties in the calculation is the unavailability of 
sufficiently accurate experimental Q-values in most cases. The 
photodisintegration rate is exponentially dependent on the Q-value. In some of 
the $Z=N+1$ nuclei, experimental binding energy values are either not 
available, or have very large errors.
In absence of experimental values, we have used the Q-values and 
their errors adopted in Refs. \cite{old} and \cite{audi} for the network calculation.
 TALYS code uses the Duflo-Zuker\cite{DZ,DZ1} formula
for masses for which experimental values are not available. These have been 
kept unchanged as the small difference between the values predicted by the 
formula and the values adopted in \cite{old} and \cite{audi} do not 
significantly affect the reaction rates. 
The Q-values used for the calculation are indicated in table 1. 

The measured half life values for beta-decay have been taken from 
the compilation by Audi {\em et al.}\cite{Audi1} except in the case of
the case of $^{65}$As. For this nucleus, a measured value of 0.128
second from Ref. \cite{as65} has been assumed.   
If measured values are not available, we have adopted the values from the 
work by M\"{o}ller {\em et al.}\cite{Moller}.

\section{Results}

This section is subdivided in three parts. In the first part, we discuss
the results of RMF calculations very briefly. We concentrate on the 
results on the nuclei where experimental measurements on 
($p,\gamma$) reactions have been performed. In the second part, we compare the 
available reaction data with the microscopic calculations obtained in 
the procedure described above. In the third subsection, we discuss the 
proton capture reactions at the waiting points.

\subsection{Relativistic mean field calculations}

We present a very brief summary of the results for only those nuclei for which 
experimental measurements of ($p,\gamma$) reaction are available. In table 2 
we compare the theoretical results for binding energy and charge radius with 
experimental information wherever available. Charge radius has been chosen 
for comparison as it provides the simplest measure of the charge density 
distribution in the nucleus. Some of the binding energy results were 
presented in Ref. \cite{NiCu} earlier. The binding energy values from the mean 
field approach have been corrected using the formalism developed in 
Ref. \cite{rmfcor,rmfcor1}. The experimental binding energy and charge radius 
values are from Refs. \cite{audi} and \cite{radius}, respectively. 

\begin{figure}[htb]
\resizebox{16cm}{!}{\includegraphics{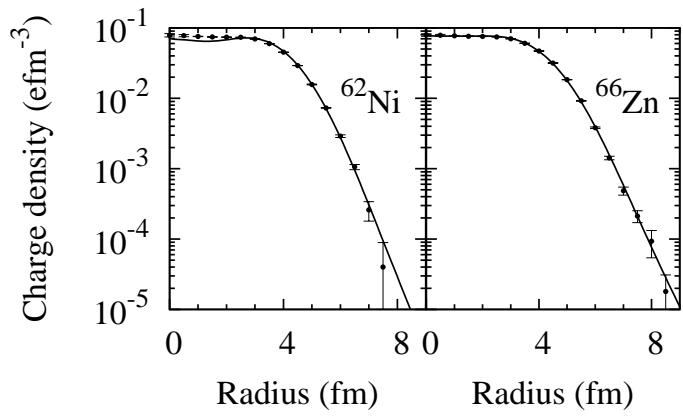}}
\caption{\label{chden} Calculated charge density in $^{62}$Ni and $^{66}$Zn (solid lines)
compared with experimental measurements (filled circles) from Ref. \cite{chden}.}
\end{figure}

Charge radius has been calculated from the charge density of the nucleus.
The charge density, in  turn, has been obtained from the point proton
density $\rho_p$ by taking into account the finite size of the proton.
The point proton density is convoluted with a Gaussian form factor
$g({\bf r})$,

\begin{eqnarray}
\rho_{ch}(\mathbf{r}) = \int e\rho_p(\mathbf{r'})g(\mathbf{r}-\mathbf{r'})d\mathbf{r'}\\
g(\mathbf{r}) = (a\sqrt{\pi})^{-3}\exp(-r^2/a^2) \end{eqnarray}
with $a=0.8$ fm.  We plot in Fig. \ref{chden} the calculated charge density for 
$^{62}$Ni and $^{66}$Zn as representatives of our results. Experimental 
measurements from Wohlfahrt {\em et al.}\cite{chden} are shown as filled 
circles. One can see that the theoretical and experimental values agree very
well, particularly at larger radii values, which is the region expected to
contribute to the optical potential at low projectile energy. Other nuclei
also show similar agreement.

One can see that the charge radius values are also reproduced to considerable 
accuracy. There were two obvious typographical errors in the theoretical
charge radii values of $^{62,64}$Ni in Ref. \cite{NiCu}.  One can see that the 
experimental quantities are adequately described in the present formalism.

\subsection{S-factors of ($p,\gamma$) reactions}

Cross-sections of low-energy $(p,\gamma)$ reactions, for which experimental 
data are available in the $A=60-80$ region, have been calculated.
The energy relevant to the $rp$-process in this mass region lies between
1.1 to 3.6 MeV. As the cross-section  varies very rapidly at such low energy, 
a comparison between theory and experiment is rather difficult. The usual 
practice in low energy nuclear reaction is to compare another key observable, 
{\em viz.} S-factor. It is given by 
\begin{equation}
S(E)=E\sigma(E)e^{2\pi\eta}
\end{equation}
where E is the energy in centre of mass frame in KeV, $\sigma(E)$ indicates reaction
cross-section in barn and $\eta$ is
 the Sommerfeld parameter with  
\begin{equation}
 2\pi\eta=31.29 Z_{p}Z_{t}\sqrt{\frac{\mu}{E}}
\end{equation}
Here, $Z_{p}$ and $Z_{t}$ are the charge numbers of the projectile and the
target, respectively and 
$\mu$ is the reduced mass (in amu). The quantity S-factor varies much more slowly than reaction  
cross-sections as the exponential energy dependence of cross-section
is not present in it. For this reason, we calculate this quantity
and compare it with experimentally extracted values.

As already pointed out, our calculations, being more microscopic, are 
more  restricting. Yet, the rate  depends on the models of the level density 
and the E1 gamma strength function adopted in the calculation of cross sections.
Phenomenological models are usually fine tuned for nuclei near the
stability valley. Microscopic prescriptions, on the other hand, can
be extended to the drip lines, which is a requirement for  
calculating the rates at the waiting points. Hence, microscopic approach  
has been assumed for all nuclei. 
We have calculated our results with microscopic level densities
in Hartree-Fock (HF) and Hartree-Fock-Bogoliubov (HFB) methods, calculated 
in TALYS by Goriley and Hilaire, respectively. For E1 gamma strength 
functions, results derived from 
HF+BCS and HFB calculations have been employed. 
All these options are available in the TALYS data base.

The real part of the potential has been obtained by normalizing the folded 
DDM3Y potential by a factor of 0.7, while the imaginary part, by a factor of 
0.1, so as to explain the S-factors obtained in the above experiments.
In Ref. \cite{NiCu}, we used a slightly different normalization.
There, for a good fit, we also used another parameter, $G_{norm}$, for
normalizing the gamma-strengths. However, we found that this parameter varies 
from nucleus to nucleus. If the present calculation has to be extended to
unknown nuclei, the approach is clearly inadequate. To overcome this difficulty, 
we have used a different normalization in the present work as stated above.

In Figs. 2-4, we plot the results of our calculations against 
S-factors extracted from experimental cross section values. 
The target nuclei are indicated in the figures.
In $^{62,64}$Ni, the experimental values are from Tingwell
{\em et al.}\cite{ni62} and Sevior {\em et al.}\cite{Cu65}, respectively.
For $^{63,65}$Cu, the experimental results are from Ref.\cite{Cu65} (empty
 circles) and the Ph.D. thesis of Qiang\cite{cu63}  (filled circles).
In Fig. \ref{znsf}, the experimental values for $^{64}$Zn are extracted from
Refs.\cite{Se741} (empty circles), \cite{zn64} (filled circles) and \cite{zn641}
(empty triangles). The results for $^{66}$Zn are from Skakun {\em et al.}
\cite{zn641} while those from $^{67,68}$Zn are from Refs.
\cite{Se741,zn68},  respectively. The results for $^{70}$Ge are taken from
Kiss {\em et al.}\cite{Ge70}. S-factors for $^{76,77}$Se are from Gyurky
{\em et al.}\cite{Se742} and Krivonosov {\em et al.}\cite{Se741}, respectively. S-factors
for $^{74}$Se$(p,\gamma)$ reactions are also from Refs.\cite{Se741}(empty circles) and \cite{Se742}
(filled circles).
 In some instances, the actual numerical values have been 
obtained from the
website of the National Nuclear Data Center\cite{nndc} of the Brookhaven 
National Laboratory.

We find that, except in the case of $^{77}$Se the results for HF plus BCS and 
HFB calculations describe the S-factors reasonably well. They both reproduce
the general trend and, except for a few isolated points, come very close to
experimental values. In many of the above measurements, the 
errors were not given. Some of the measurements are also old. Taking all these 
facts in to consideration, it is easy to see that the HFB method scores 
particularly well. Interestingly, more recent experimental data, whenever
available, show better agreements with HFB results
than previous measurements. The HFB approach also is known to work well for 
nuclei away from the stability valley. So, for later calculations, we have 
employed the level density and E1 gamma strength values from HFB calculation.
The normalization of the potential, as described above, has also been assumed.

\subsection{Waiting point nuclei in $rp$ process}

Effective half life values for seed nuclei $^{64}$Ge, $^{68}$Se,
$^{72}$Kr, and $^{76}$Sr are shown in Figs. 5-8 using continuous lines. For 
comparison, we have also plotted the results calculated from the rates in 
Rauscher
{\em et al.}\cite{astro} by dash-dotted lines. 
The effects of the uncertainties on the half life values in the Q-values have 
been indicated in the figures by dotted lines.

One can see that the effective half life of the waiting point nuclei indeed 
gets reduced by two proton capture, reaching the minimum around 1-2 GK temperature. However, the rate 
of change is different at different waiting points. We need to check whether 
the reduction is sufficient for nucleosynthesis to bridge the waiting points
and to proceed along the $Z=N$ line.

Fig. \ref{ge64} shows the change in the 
effective half life of $^{64}$Ge in explosive hydrogen rich environment. The 
half life decreases and possibly goes  to a value substantially less 
than ten seconds, the minimum duration of an X-ray burst. However, one sees 
that the uncertainty in mass measurement prevents one from reaching any firm conclusion. 
Depending on the actual value of the masses, it may even be possible that a 
burst of the order of ten seconds cannot bridge this waiting point effectively. 

In Figs. \ref{se68} and \ref{kr72} one can see that effective half lives of 
$^{68}$Se and $^{72}$Kr  are not affected significantly by rapid proton 
processes. This conclusion remains unchanged even if the errors in Q-values are 
taken into account. Thus, the most likely scenario is that the $rp$-process path 
shifts to more stable nuclei and proceed along a different path. Possibilities
of bridging these two points in astrophysical environment have been briefly
discussed later in this work. In the case of
$^{76}$Sr, on the other hand, the half life decreases by more than a factor of 
two.
It is clear that this particular waiting point is easily bridged by 
$rp$-process.

Is it possible for the waiting points at $^{68}$Se and $^{72}$Kr to be bridged 
in environmental conditions other than assumed in the calculation?
If the two proton capture rate exceeds that of beta-decay at the waiting point, 
we may consider it to be effectively bridged by rapid proton capture. In Fig. 
9 we show the densities above which the waiting points at $^{64}$Ge, $^{68}$Se, 
$^{72}$Kr, and $^{76}$Se are bridged, as a function of temperature. We see that for a density 
of $10^6$gm/cm$^3$, the first of the above waiting points are bridged
over a large temperature range. The waiting point at $^{76}$Sr has a narrower 
temperature window but can definitely be bridged at the density. The point
at $^{68}$Se requires slightly higher density which may perhaps be available 
in astronomical environments. However, the waiting point at $^{72}$Kr presents 
a completely different picture. One sees that a density in excess of 
$10^7$ gm/cm$^3$ is required for effectively bridging this waiting point. Such 
a high density is not expected even in X-Ray burster environment. Thus this 
waiting point is expected to stall the $rp$-process nucleosynthesis and 
shift it towards the stability valley.

We note that the rates from Rauscher
{\em et al.}\cite{astro} produce nearly 
identical results except in the case of $^{72}$Kr. The rates used in the 
present calculation are slightly smaller, a fact that gives rise to a slight
decrease in the temperature where the  half life has a minimum. As our 
calculation is microscopic in nature, while at the same time depending on 
local density approximation, we expect the results obtained
in the present work to be more reliable. In all the 
cases studied above, including that at $^{72}$Kr, the conclusions arrived at 
the present work do not change to any appreciable extent in case the rates from 
Ref. \cite{astro} are adopted. This is to be expected as both the present
work and Ref. \cite{astro} use the Hauser-Feshbach formalism.

\section{Summary}

In summary, RMF calculations has been performed in various nuclei in
 mass 60-80 region using FSU Gold Lagrangian density.
Cross sections for $(p,\gamma)$ reactions in nuclei with 
$60<A<80$ have been calculated using a microscopic optical potential obtained 
by folding the DDM3Y effective interaction with the theoretical nuclear 
densities. The parameters employed in the microscopic optical potential have 
been fixed after comparison of experimental S-factors with calculated ones. 
We have found that the level density and E1 gamma strength obtained form 
HFB calculation are eminently suited to describe the observed S-factors.
Astrophysical rates for proton capture and 
photodisintegration  rates have been calculated in $rp$-process waiting point
nuclei with $60<A<80$ using the same approach.  The possibility of bridging the 
waiting points by two 
proton capture has been investigated.  We see that unlike the other waiting 
points studied, the one at $^{72}$Kr is unlikely to be bridged by two proton 
capture.

\section*{Acknowledgments}

This work has been carried out with financial assistance of the UGC sponsored
DRS Programme of the Department of Physics of the University of Calcutta. 
Chirashree Lahiri acknowledges the grant of a fellowship awarded by the UGC.

\newpage
\begin{table}[h]
\begin{center}
\caption{Q-values adopted in the calculation}
\begin{tabular}{cl}
 \hline
Reaction & \multicolumn{1}{c}{Q-values(MeV)}\\ \hline
$^{64}$Ge$(p,\gamma)^{65}$As & -0.255$\pm$ 0.104 \cite{prc}\\ 
$^{65}$As$(p,\gamma)^{66}$Se & ~2.350 $\pm$ 0.200 \cite{prc}\\ 
$^{68}$Se$(p,\gamma)^{69}$Br & -0.785 $^{+0.034}_{-0.040}$ \cite{brmass}\\ 
$^{69}$Br$(p,\gamma)^{70}$Kr & ~2.605$\pm$ 0.16 \cite{old}\\ 
$^{72}$Kr$(p,\gamma)^{73}$Rb & -0.71 $\pm$ 0.10 \cite{prl}\\ 
$^{73}$Rb$(p,\gamma)^{74}$Sr & ~2.18 $\pm$ 0.14 \cite{prl}\\ 
$^{76}$Sr$(p,\gamma)^{77}$Y & -0.050 $\pm$ 0.072 \cite{audi}\\ 
$^{77}$Y$(p,\gamma)^{78}$Zr & ~2.087 $\pm$ 0.507 \cite{audi}\\\hline
\end{tabular} 
\end{center}
\end{table}
\begin{table}[h]
\begin{center}
\caption{B.E. and radius values in selected nuclei}
\begin{tabular}{ccccc}
 \hline
Nucleus & \multicolumn{2}{c}{Binding Energy (MeV)}&\multicolumn{2}{c}
{Charge radius ($fm$)}\\
&  \multicolumn{1}{c}{Theo.}& \multicolumn{1}{c}{ Exp.} & \multicolumn{1}{c}{Theo.} &\multicolumn{1}{c}{ Exp.}\\\hline
$^{62}$Ni&544.71&545.26&3.854&3.841\\
$^{64}$Ni&561.97&561.76&3.868&3.859\\
$^{63}$Cu&551.17&551.38&3.889&3.883\\
$^{65}$Cu&569.43&569.21&3.908&3.902\\
$^{64}$Zn&558.815&559.098&3.923&3.929\\
$^{66}$Zn&578.139&578.136&3.937&3.950\\
$^{67}$Zn&585.985&585.188&3.943&\\
$^{68}$Zn&595.94&595.386&3.950&3.966\\
$^{70}$Ge&611.353&610.520&4.021&4.041\\
$^{74}$Se&642.890&642.890&4.103&4.070\\
$^{76}$Se&662.234&662.072&4.110&4.140\\
$^{77}$Se&669.964&669.491&4.113&4.140\\
\hline
\end{tabular}
\end{center}
\end{table}

\begin{figure*}[htb]
\resizebox{18cm}{!}{\includegraphics{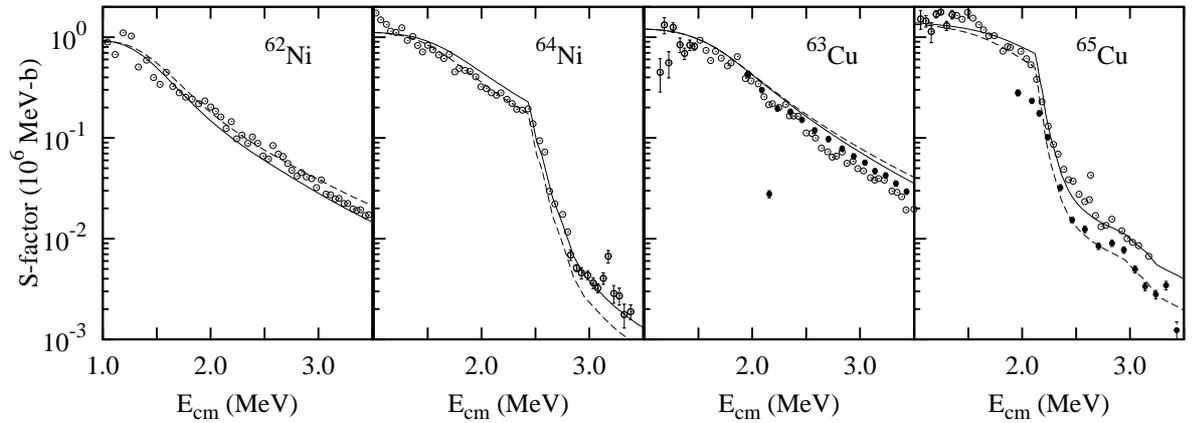}}
\caption{\label{nicusf}S-factors extracted from experimental measurements
compared with theory for $^{60,62}$Ni and $^{63,65}$Cu. Solid and dashed lines indicate respectively 
the results of the HF+BCS
and HFB approaches for level density and E1 gamma strength. $E_{cm}$  is centre of mass frame energy. 
}
\end{figure*} 

\begin{figure*}[hbt]
\resizebox{18cm}{!}{\includegraphics{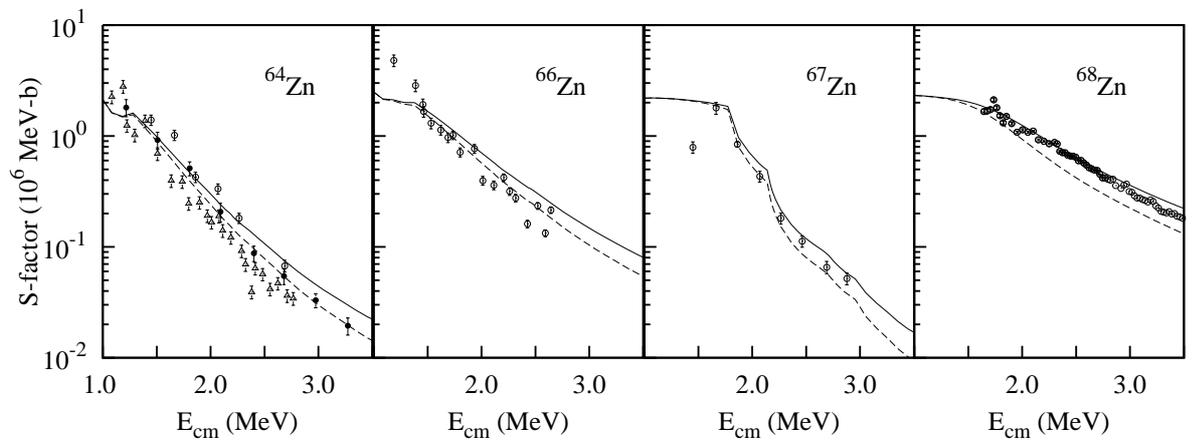}}
\caption{\label{znsf}S-factors extracted from experimental measurements
compared with theory for $^{64,66,67,68}$Zn. See caption of Fig. \ref{nicusf}
for details.
}
\end{figure*}

\begin{figure*}[htb]
\resizebox{18cm}{!}{\includegraphics{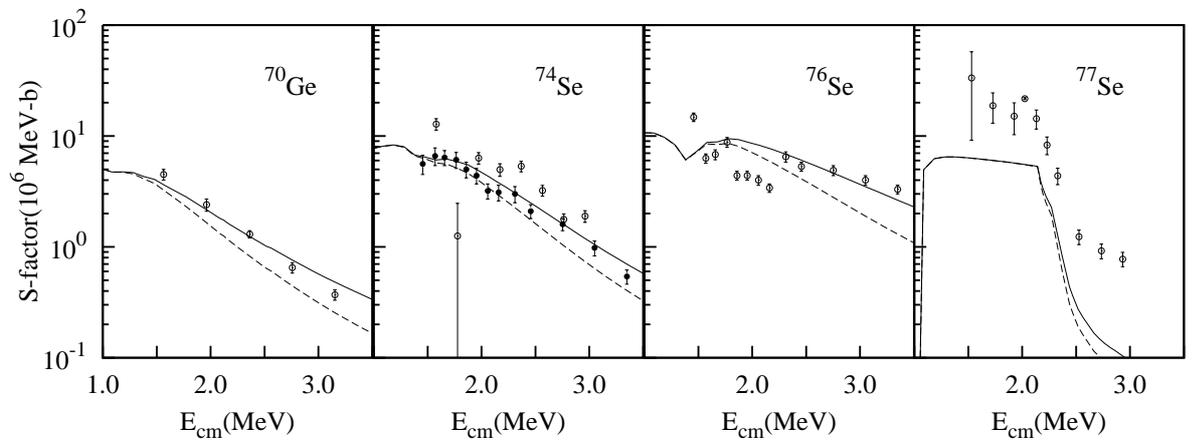}}
\caption{\label{gesesf}S-factors extracted from experimental measurements
compared with theory for $^{70}$Ge and $^{74,76,77}$Se. 
 See caption of Fig. \ref{nicusf}
for details.
}
\end{figure*} 

\begin{figure}[bht]
\resizebox{\columnwidth}{!}{\includegraphics{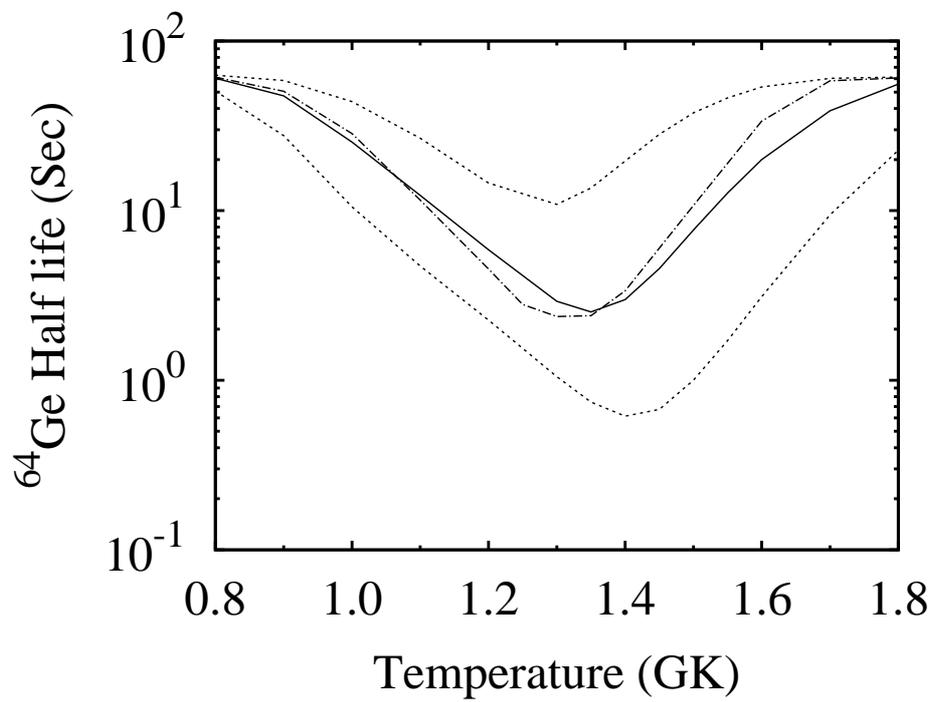}}
\caption{\label{ge64} Effective half life values of $^{64}$Ge as a function of temperature. 
The solid line represents the results of our calculation while 
the dashed lines mark the two extremes for the errors in the Q-values of the 
reactions involved. The dash dotted line shows the results obtained using the rates 
from \cite{astro}. }
\end{figure}

\begin{figure}[ht]
\resizebox{\columnwidth}{!}{\includegraphics{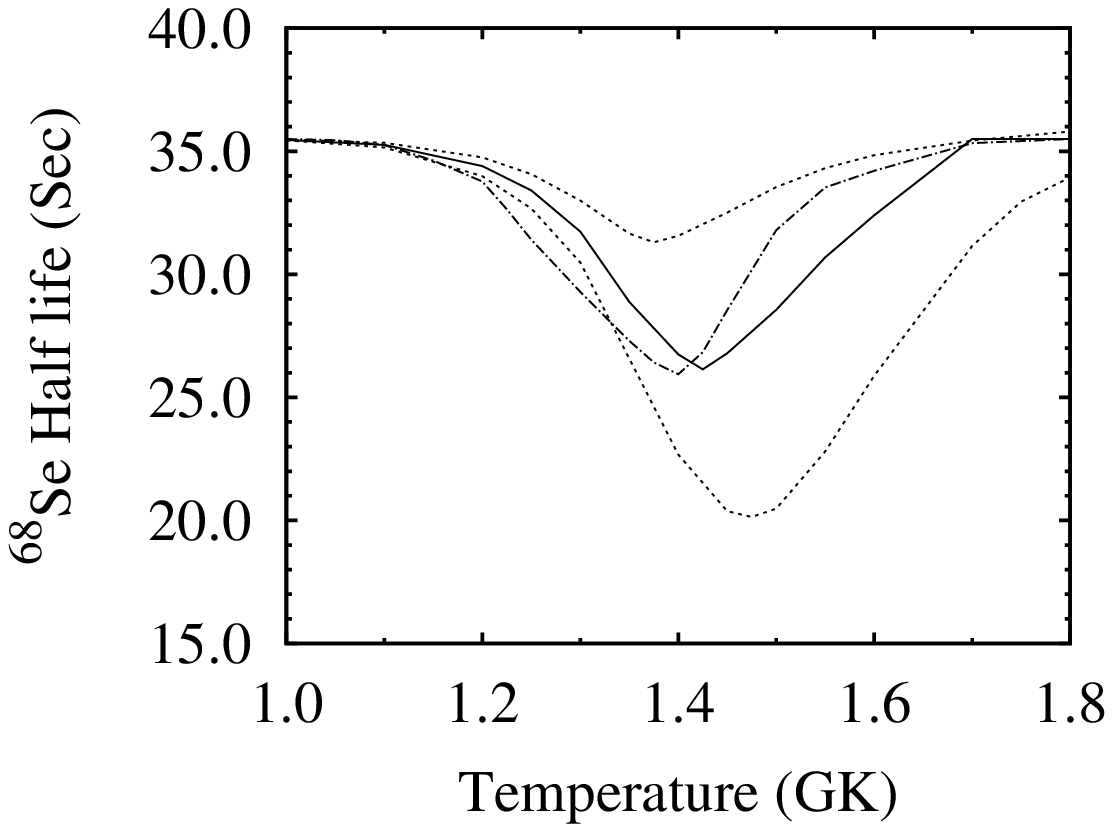}}
\caption{\label{se68} Effective half life values of $^{68}$Se as a function of temperature. See caption of 
Fig. \ref{ge64} for details.}
\end{figure}

\begin{figure}[ht]
\resizebox{\columnwidth}{!}{\includegraphics{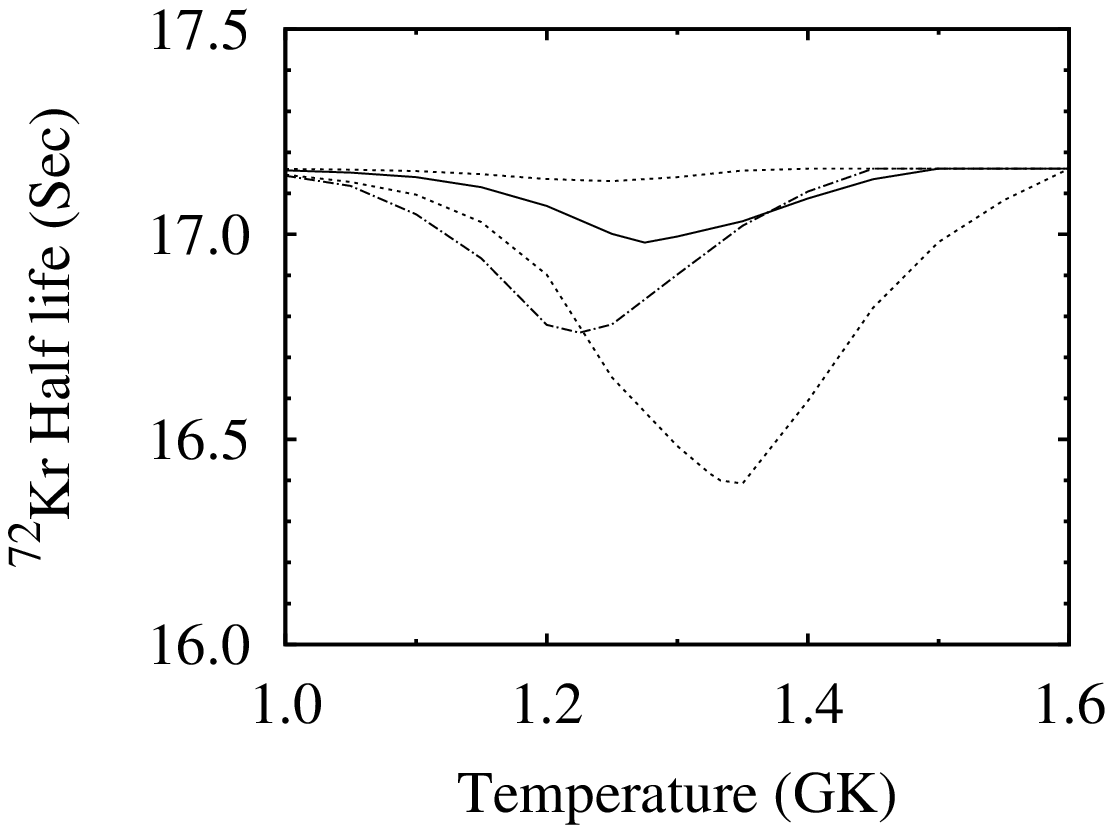}}
\caption{\label{kr72} Effective half life values of $^{72}$Kr as a function of temperature. See caption of 
Fig. \ref{ge64} for details.}
\end{figure}

\begin{figure}[ht]
\resizebox{\columnwidth}{!}{\includegraphics{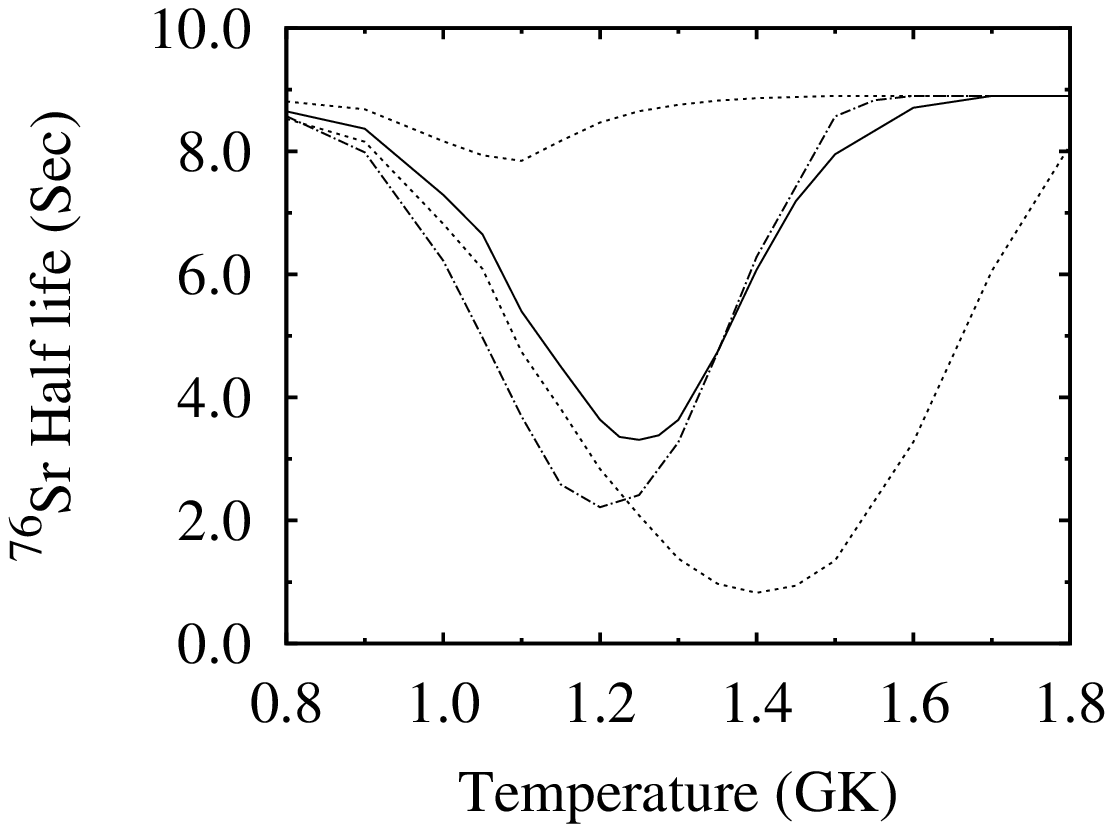}}
\caption{\label{sr76} Effective half life values of $^{76}$Sr as a function of temperature. See caption of 
Fig. \ref{ge64} for details.}
\end{figure}

\begin{figure}[ht]
\resizebox{\columnwidth}{!}{\includegraphics{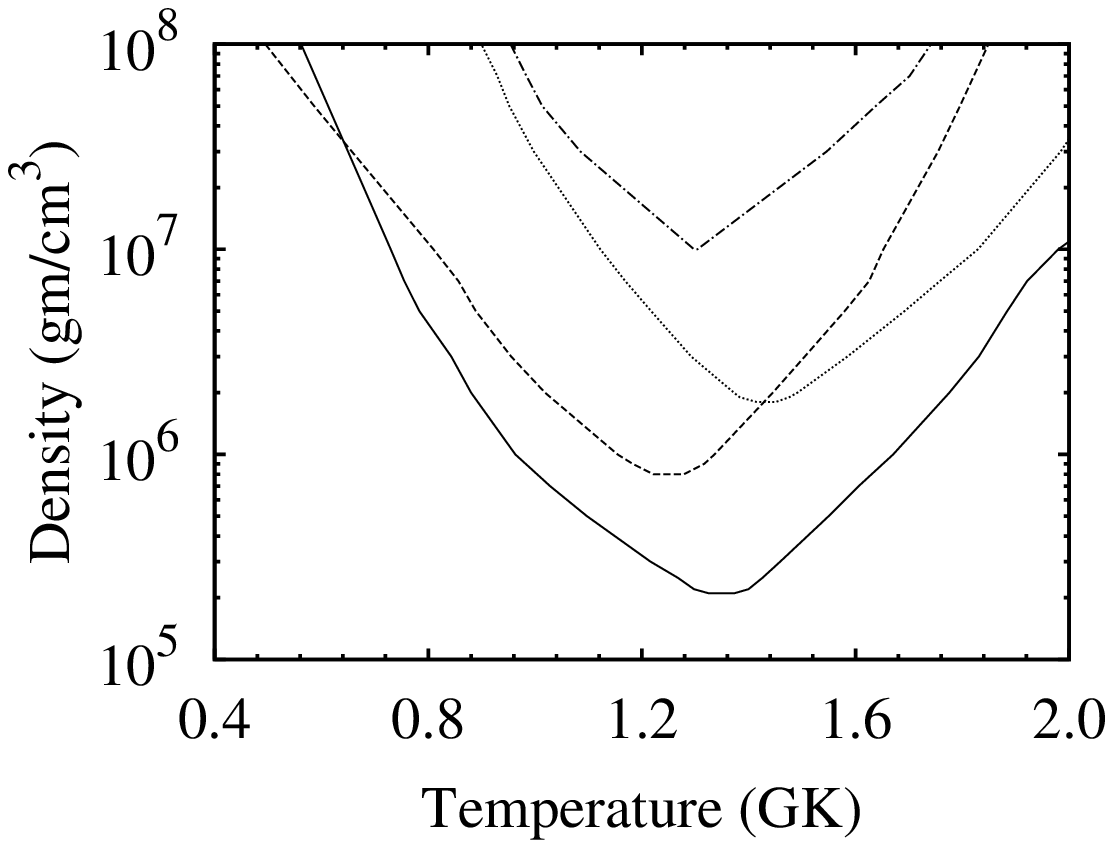}}
\caption{\label{dens} Densities above which the various
waiting points are found to be to be effectively bridged at different 
temperatures. The waiting points are $^{64}$Ge (continuous curve), $^{68}$Se
 (dotted curve), $^{72}$Kr (dash-dotted curve) and $^{76}$Sr (dashed curve).} 
\end{figure}

\begin{thebibliography}{99}
\bibitem{RMF1} See {\em e.g.} P. Ring, Prog. Part. Nucl. Phys. {\bf 37} (1996)
193.
\bibitem{decay}M. Bhattacharya and G. Gangopadhyay, Phys. Rev. C {\bf 77}  (2008) 047302.
\bibitem{NiCu}G. Gangopadhyay, Phys. Rev. C{\bf 82}  (2010) 027603.
\bibitem{CBe}G. Gangopadhyay and S. Roy, J. Phys. G: Part. Nucl. Phys {\bf 31}
(2005) 1111.
\bibitem{mom}E. Bauge, J.P. Delaroche and M. Girod, Phys. Rev. C {\bf 63} (2001)
024607.
\bibitem{book}C. Illiadis, \textit{Nuclear Physics of the Stars}
(Wiley-VCH Verlag GmbH, Weinheim 2007).
\bibitem{rep}See for example, H. Schatz {\em et al.}, Phys. Rep. {\bf 294} (1998) 167.
\bibitem{astro}T. Rauscher and F.K. Thielemann, At. Data Nucl. Data Tabl. {\bf 75} (2000) 1.
\bibitem{rate1}T. Rauscher and F.K. Thielemann, At. Data Nucl. Data Tabl. {\bf 79} (2000) 47.
\bibitem{old}H. Schatz, Int. J. Mass Spec. {\bf 251} (2006) 293.
\bibitem{fsu}B.G. Todd-Rutel and J. Piekarewicz, Phys. Rev. Lett.{\bf 95} (2005) 122501.
\bibitem{MB}M. Bhattacharya and G. Gangopadhyay, Phys. Rev. C {\bf 72}  (2005) 
044318.
\bibitem{MB1}M. Bhattacharya and G. Gangopadhyay, Fizika (Zagreb) {\bf 16} (2007) 113.
\bibitem{ddm3y1} A.M. Kobos, B.A. Brown, R. Lindsay and G. R. Satchler,
Nucl. Phys. {\bf A425}  (1984) 205.
\bibitem{ddm3y2} A.K. Chaudhuri, Nucl. Phys. {\bf A449}  (1986) 243. 
\bibitem{ddm3y3}A.K. Chaudhuri, Nucl. Phys.{\bf A459}  (1986) 417.
\bibitem{ddm3y}See {\em e.g.} D. N. Basu,  Journal of Physics G {\bf 30}  (2004) B7.
\bibitem{SO} R.R. Scheerbaum,  Nucl. Phys. {\bf A257}  (1976) 77.
\bibitem{talys}A.J. Koning {\em et al.}, Proc. 
Int. Conf.  Nucl. Data  Science  Tech., 
April 22-27, 2007, Nice, France, EDP Sciences, (2008) p. 211.
\bibitem{audi}G.Audi, A.H.Wapstra and C.Thibault, Nucl. Phys. {\bf A729} (2003) 337.

\bibitem{DZ}J. Duflo and A.P. Zuker, Phys. Rev. C {\bf 52} (1995) R23. 
\bibitem{DZ1}J. Duflo and A.P. Zuker,  Phys. Rev. C {\bf 59}  (1999)  R2347.
\bibitem{prc}P. Schury {\em et al.}, Phys. Rev. C {\bf 75}  (2007) 055801.
\bibitem{brmass}A.M. Rogers {\em et al.}, (2010) Arxiv-1009.2950.
\bibitem{prl}D. Rodr\'{i}guez {\em et al.}, Phys. Rev. Lett. {\bf 93}  
 (2004) 161104.
\bibitem{Audi1} G. Audi, O. Bersillon, J. Blachot  and A.H. Wapstra,
Nucl. Phys. {\bf A729}  (2003) 3.
\bibitem{as65}M.J. L\'{o}pez Jim\'{e}nez {\em et al.}, Phys. Rev. C {\bf 66}   
(2002) 025803.
\bibitem{Moller}P. M\"{o}ller, J. R. Nix and K. L. Kratz, Atom. Data Nucl. Data
Tabl. {\bf 66}  (1997) 131.
\bibitem{rmfcor}M. Bhattacharya  and G. Gangopadhyay, Phys. Lett. B {\bf 672}  
 (2009) 182.
\bibitem{rmfcor1}G. Gangopadhyay, J. Phys. G : Part. Nucl. Phys {\bf 37}  (2010 015108.
\bibitem{chden}H.D. Wohlfahrt, O. Schwentker, G. Fricke, H.G. Andersen and
E.B. Shera, Phys. Rev. C {\bf 22}, 264 (1980).
\bibitem{radius}I. Angeli, At. Data Nucl. Data Tables, {\bf 87}  (2004) 185.
\bibitem{ni62}C.I.W. Tingwell, V.Y. Hansper, S.G. Tims, A.F. Scott, A.J. Morton
and D.G.Sargood, Nucl. Phys {\bf A496}  (1988) 127.
\bibitem{Cu65}M.E. Sevior, L.W. Mitchell, M.R. Anderson, C.W. Tingwell and
D.G.Sargood, Aust. J. of Phys., {\bf 36}  (1983) 463.
\bibitem{cu63}S. Qiang, Ph.D. Thesis, University of Kentucky, 1990.
\bibitem{Se741}G.A. Krivonosov, O.I. Ekhichev, B.A. Nemashkalo, V.E. Storizhko
and V.K.Chirt, Izv. Akad. Nauk. SSSR, Ser. Fiz {\bf 41}  (1977) 2196. 
\bibitem{zn64}M.A. Famiano, R.S. Kodikara, B.M. Giacherio, V.G. Subramanian 
and A.Kayani, Nucl. Phys. A {\bf 802} (2008) 26.
\bibitem{zn641}E.A. Skakun, S.N. Utenkov, V.N. Bondarenko,A.V. Goncharov,
V.M. Mishchenko, V.I. Sukhostavets and K.V. Shebeko, Izv. Akad. Nauk. SSSR, 
Ser. Fiz {\bf 72} (2008) 402.
\bibitem{zn68}M.T. Esat, R.H. Spear, J.L. Zyskind, M.H. Shapiro,                W.A. Fowler and J.M. Davidson, Phys. Rev. C {\bf 23}  (1981) 1822.
\bibitem{Ge70}G.G. Kiss, Gy. Gyurky, Z. Elekes, Zs. Fulop, E. Somorjai,  T. 
Rauscher and M. Wiescher, Phys. Rev. C {\bf 76}  (2007) 055807.
\bibitem{Se742}Gy. Gyurky, Zs. Fulop, E. Somorjai, M. Kokkoris,
S. Galanopoulos, P. Demetriou, S. Harissopulos,
T. Rauscher and S. Goriely,  Phys. Rev. C {\bf 68}  (2003) 055803.
\bibitem{nndc}http://www.nndc.bnl.gov.
\end{thebibliography}
\end{document}